\documentclass{ws-procs9x6}

\begin{document}

\title{Looking Back at the First Decade of 21st-Century
High-Energy Physics
}
\author{John Ellis}

\address{Theory Division, CERN, \\
CH 1211 Geneva 23, \\
Switzerland \\
E-mail: John.Ellis@cern.ch}


\maketitle

\abstracts{
On the occasion of the Tenth Conference on String Phenomenology in 2011, I
review the dramatic progress since 2002 in experimental tests of fundamental
theoretical ideas. These include the discovery of (probably fermionic)
extra dimensions at the LHC, the discovery of dark matter particles,
observations of charged-lepton flavour violation, the debut of quantum
gravity phenomenology and the emergence of space-time from the string
soup.
}ž

\begin{center}
CERN-TH/2002-189 ~~~~~~ ~~~~~~ hep-ph/0208109
\end{center}

\section{Introduction}

The organizers of the Tenth Conference on String Phenomenology, $\cos^{10}
\phi$, have invited me to review the exciting developments in high-energy
physics that took place during the tumultuous first decade of the 21st
century. It is
amusing to look back at the quaint preoccupations of participants in the
first $\cos \phi$ meeting, back in 2002. At that time, accelerators had
established what was then regarded as the Standard Model, but offered no
clear indications what directions physics might take beyond it, and the
community was anxious for the LHC to be funded and completed. On the other
hand, non-accelerator experiments had provided evidence beyond the
Standard Model in the form of neutrino oscillations. Meanwhile, string
theorists were all doing $PP$ or branes in extra dimensions.

How different is today's panorama! The LHC has taken us triumphantly
beyond the Standard Model, and its financial travails back in 2002 have
been long forgotten. However, now we are anxious for the linear $e^+ e^-$
collider to be funded and completed. In parallel, cosmology has taken us
far beyond the petty models for physics beyond the Standard Model that we
were playing with back in 2002. Quantum gravity has now become an
experimental science, and we are probing directly the emergence of
space-time itself.

\begin{figure}[th]
\centerline{\epsfxsize=2.5in\epsfbox{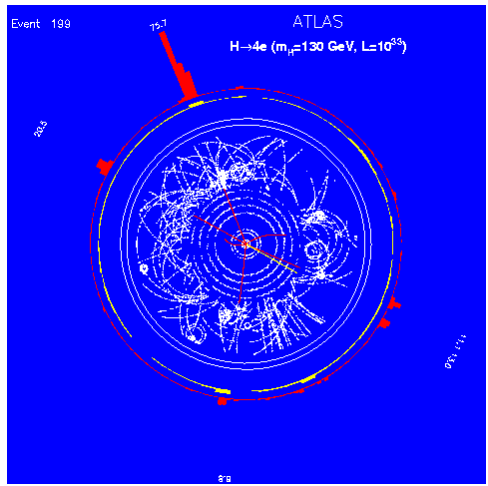}}
\caption{One of the first Higgs events observed at the LHC by the ATLAS
collaboration, in which the Higgs boson decays
into $e^+ e^- e^+ e^-$.
\label{fig:Higgs}}
\end{figure}

\begin{figure}[th]
\centerline{\epsfxsize=3.5in\epsfbox{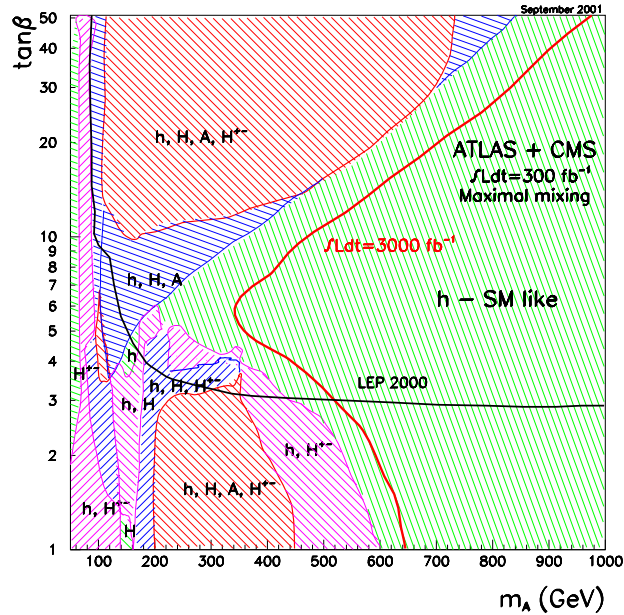}}
\caption{Illustration of the improved reach for heavier MSSM Higgs bosons
expected with the LHC upgrade to $10^{35}$~cm$^{-2}$s$^{-1}$ now
underway.
\label{fig:FG}}
\end{figure}

There is no point in reviewing here the discovery of the Higgs
boson~\cite{ATLAS+CMS}. That is an old story by now, prefaced by the
evanescent hint at LEP~\cite{LEPHiggs}, the suspense maintained by the
data from the Tevatron~\cite{CDF+D0}, and climaxed by the prompt
appearance of the Higgs boson at the LHC, as illustrated in
Fig.~\ref{fig:Higgs}. The most important remaining Higgs question is
whether the LHC will be able to discover its partners that are expected in
the Minimal Supersymmetric Standard Model (MSSM). As seen in
Fig.~\ref{fig:FG}, it is touch-and-go whether the LHC will be to find
these heavier MSSM Higgs bosons, though the LHC luminosity upgrade to
$10^{35}$ cm$^{-2}$s$^{-1}$ currently underway will certainly improve the
chances~\cite{upgrade}.

Rather than the old Higgs story, here I focus on the most important 
discovery of the LHC to date, namely ....

\section{Discovery of Extra Dimensions}

\begin{figure}[th]
\vspace*{5mm}
\centerline{\epsfxsize=2in\epsfbox{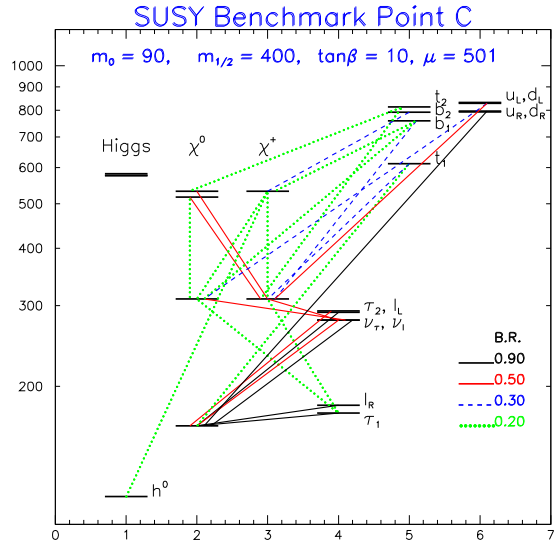}}
\vspace*{9mm}
\caption{
Typical example of the cascades of sparticle decays expected in a
scenario with fermionic extra dimensions at the LHC, which is
to be
distinguished from the type of cascade expected in a scenario with bosonic
extra dimensions, as illustrated in Fig.~\ref{fig:KKcascade}.
\label{fig:cascade}}
\end{figure}

\begin{figure}[th]
\centerline{\epsfxsize=2.5in\epsfbox{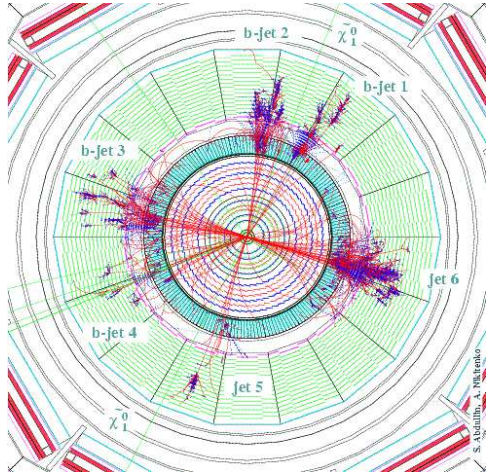}}
\caption{One of the first supersymmetric candidate events observed at the
LHC by the CMS collaboration. This event includes two energetic jets,
three leptons and missing energy.
\label{fig:susy}}
\end{figure}

\begin{figure}[th]
\centerline{\epsfxsize=2.5in\epsfbox{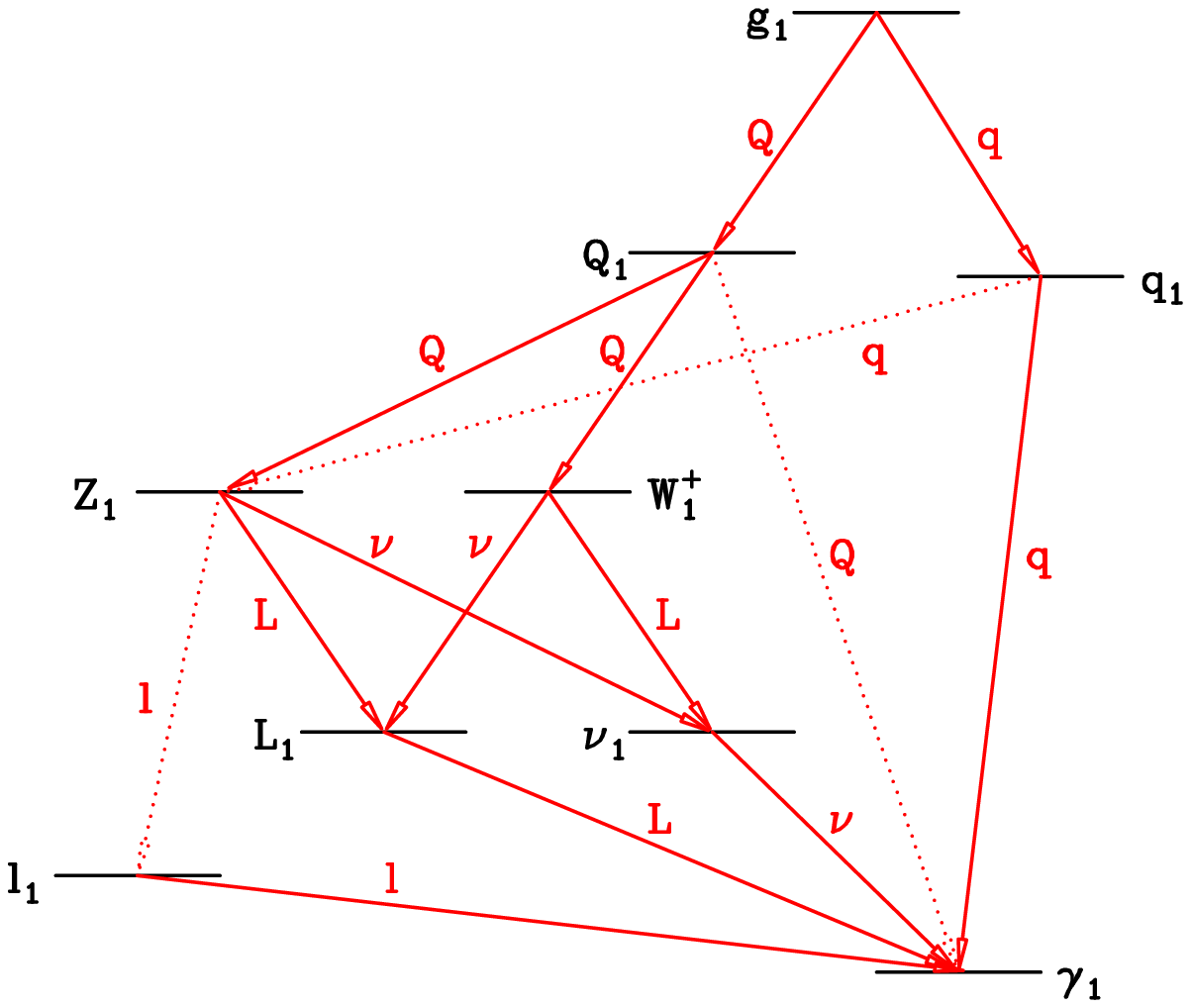}}
\caption{
Typical example of the cascades of Kaluza-Klein excitation decays expected
in a scenario with universal bosonic extra dimensions at the LHC, which is
to be distinguished from the type of cascade expected in a scenario with
fermionic extra dimensions, as illustrated in Fig.~\ref{fig:cascade}.
\label{fig:KKcascade}}
\end{figure}

The issue was never really whether the LHC would discover extra
dimensions, but whether they would be bosonic: $x_{5, 6, ...}$ or
fermionic: $\theta, {\overline \theta}$. Long before the LHC, it was
pointed out that one of the most promising sets of signals for
supersymmetry would be events with missing transverse energy, accompanied
by hadronic jets and/or leptons produced in the cascade decays of heavier
sparticles, as seen in Fig.~\ref{fig:cascade}~\cite{bench}. Events
resembling these predictions were indeed found almost as soon as the LHC
switched on, see for example Fig.~\ref{fig:susy}~\cite{ATLAS+CMS}.
However, already back in 2002, it had been pointed out that universal
extra-dimensional models with conserved Kaluza-Klein parity also predict
missing-energy events. Just like the sparticle cascades, the sequential
decays of Kaluza-Klein states produce accompanying jets and leptons, as
seen in Fig.~\ref{fig:KKcascade}~\cite{KK}.

\begin{figure}[th]
\centerline{\epsfxsize=3in\epsfbox{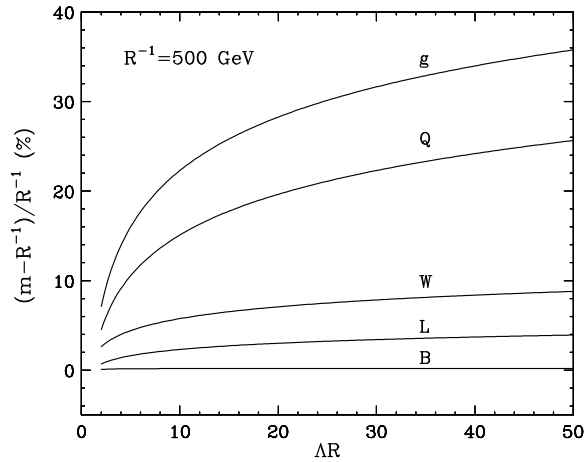}}
\caption{
Sample calculation of the renormalization of the masses of Kaluza-Klein
excitations in a model with universal extra dimensions.
\label{fig:KKmasses}}
\end{figure}

Of course, distinguishing the supersymmetric and Kaluza-Klein scenarios
turned out to be relatively simple, in principle. For similar masses, the
cross sections at the LHC were different, as well as their angular
distributions.  Moreover, the mass differences between different
Kaluza-Klein states were relatively small, as seen in
Fig.~\ref{fig:KKmasses}~\cite{KK}, because of the small
renormalization-group range compared with the GUT scale anticipated in
supersymmetry. Using these clues, the supersymmetric interpretation of the
LHC missing-energy events has gained the upper hand, though there may
still be some die-hard advocates of the Kaluza-Klein interpretation.

\begin{figure}[th]
\centerline{\epsfxsize=3.5in\epsfbox{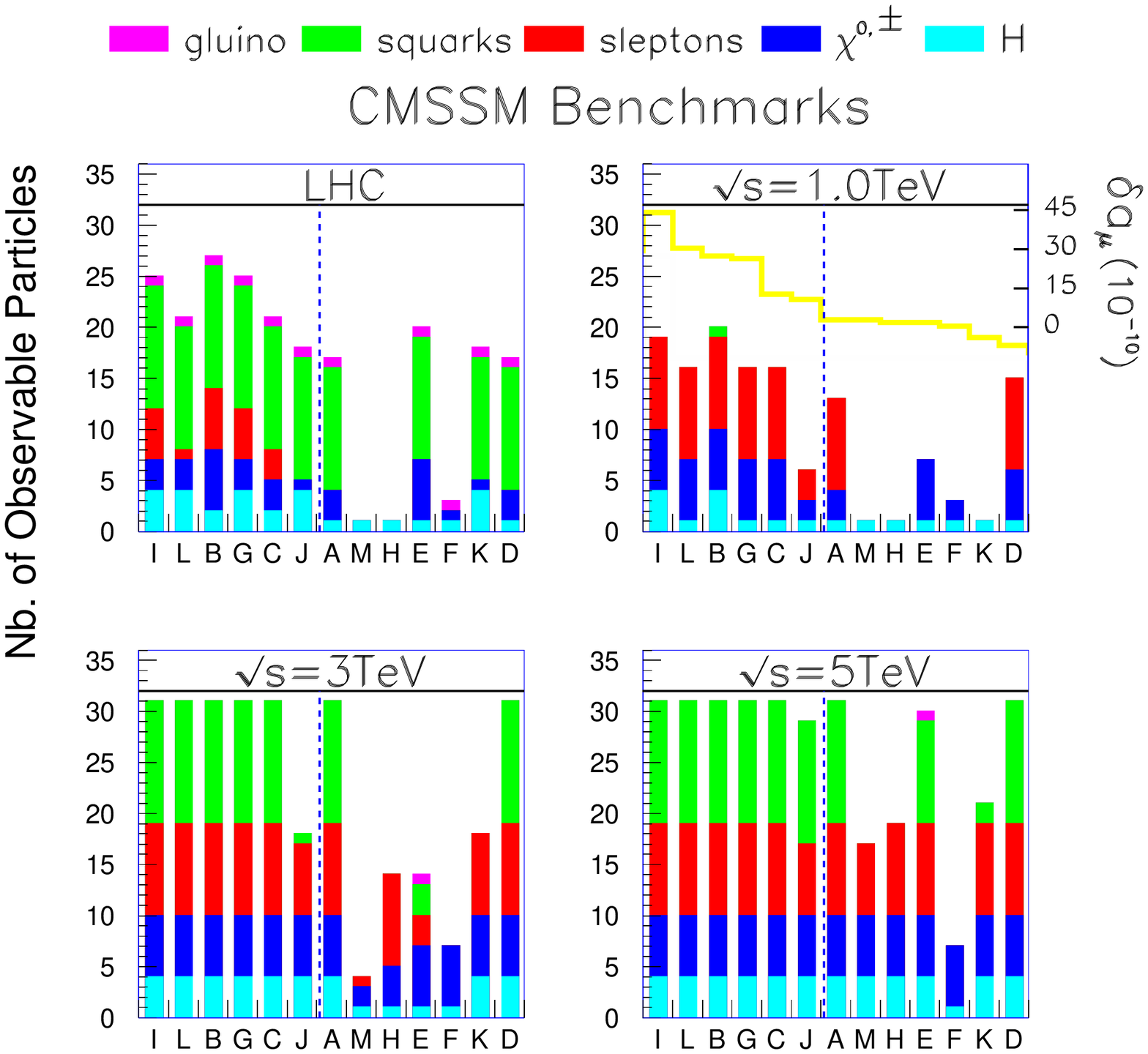}}
\caption{
Examples of the numbers of different states in scenarios with extra
fermionic dimensions that are observable at different
colliders.
\label{fig:Manhattan}}
\end{figure}

\begin{figure}[th]
\centerline{\epsfxsize=2.5in\epsfbox{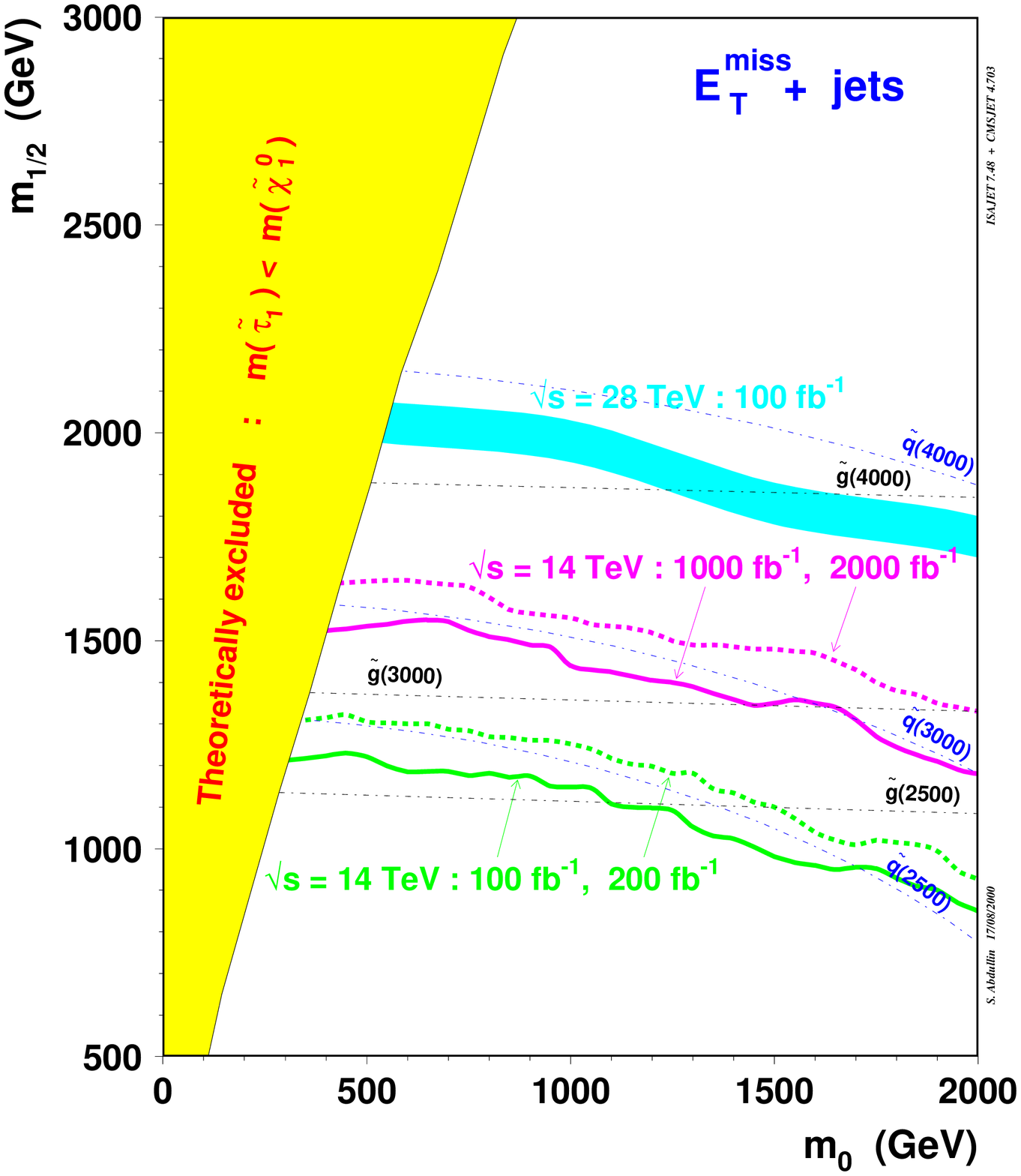}}
\caption{
The LHC luminosity upgrade to $10^{35}$~cm$^{-2}$s$^{-1}$ will enable the
physics reach for sparticles and extra bosonic dimensions to be extended
significantly.
\label{fig:upgrade}}
\end{figure}

The discovery of supersymmetry at the LHC certainly did not come as a
surprise, since most of the allowed parameter space was known to be
accessible to the LHC, as foreshadowed in
Fig.~\ref{fig:Manhattan}~\cite{bench}. However, the initial LHC
configuration has not been able to measure all the sparticle masses and
other properties one should like to know, providing a second motivation
for the LHC luminosity upgrade, as seen in
Fig.~\ref{fig:upgrade}~\cite{upgrade}.

The community has long known that the TeV-scale linear collider now under
construction would be needed to unravel sparticle (or Kaluza-Klein)
spectroscopy, which is why we have been so anxious about its financing.
Just as the corresponding LHC problems were overcome in 2002, we have been
relieved this year that the analogous linear collider issues have also
been resolved. We look forward soon to detailed measurements of Higgs
decay branching ratios, which may tell us indirectly whether the heavier
supersymmetric Higgs bosons are really there, and to better measurements
of the sparticle (or Kaluza-Klein) masses. These will finally lay to rest
the debate between gravity-mediated and other scenarios for supersymmetry
breaking, and perhaps even finally convince Kaluza-Klein die-hards that
the extra dimensions are indeed fermionic.

\begin{figure}[th]
\centerline{\rotatebox{-90}{\epsfxsize=2.5in\epsfbox{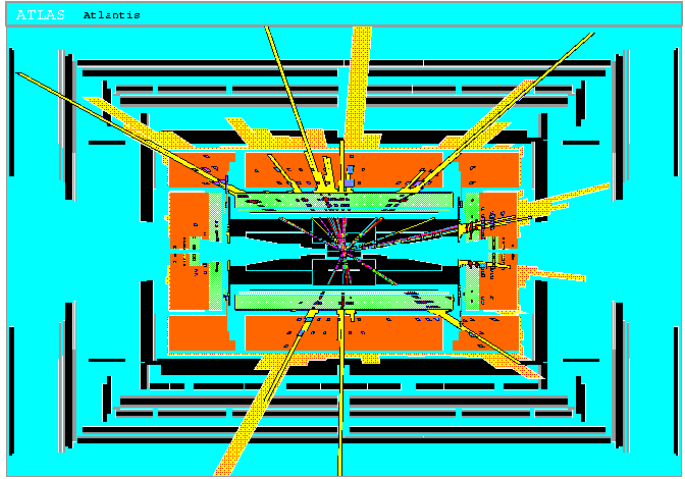}}}
\caption{Candidate for a black-hole production event recently observed at
the LHC.
\label{fig:BH}}
\end{figure}

Their last-ditch stand has, however, been encouraged by the remarkable LHC
event reported last year, shown in Fig.~\ref{fig:BH}, which has all the
characteristics expected of TeV-scale black-hole production~\cite{BH}. It
has many jets, predominating over leptons and missing energy. Time and the
LHC luminosity upgrade will tell us whether this was a statistical
fluctuation or a harbinger of Hawking radiation.

\section{Discovery of Dark Matter}

\begin{figure}[th]
\centerline{\epsfxsize=3in\epsfbox{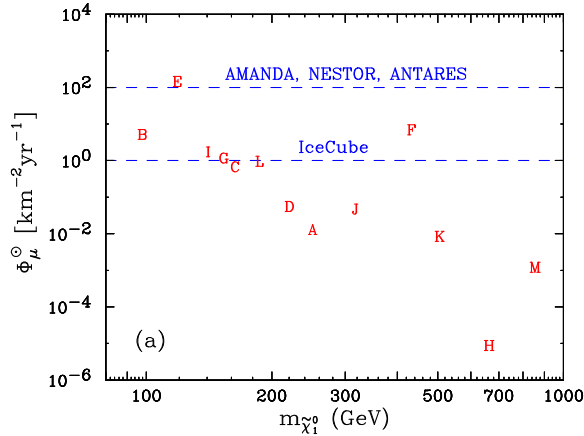}}
\caption{
The recent discovery by the IceCube Collaboration of of muons due to
energetic neutrinos from the Sun can easily be accommodated in a number of
scenarios with extra fermionic dimensions.
\label{fig:Sun}}
\end{figure}

It was known at the beginning of the past decade that astroparticle
experiments would have a fair chance of detecting supersymmetric dark
matter, if it existed. The direct search for the scattering of dark-matter
particles in the laboratory has not yet been successful, despite the best
efforts of the CDMS collaboration in the Soudan mine~\cite{CDMS}. However,
the recent observations by IceCube~\cite{IceCube} of energetic muons from
the direction of the Sun, believed to originate from the interactions of
energetic solar neutrinos, are compatible with several of the benchmark
supersymmetric scenarios proposed a decade ago, as seen in
Fig.~\ref{fig:Sun}~\cite{EFFMO}.

\begin{figure}[th]
\centerline{\epsfxsize=3in\epsfbox{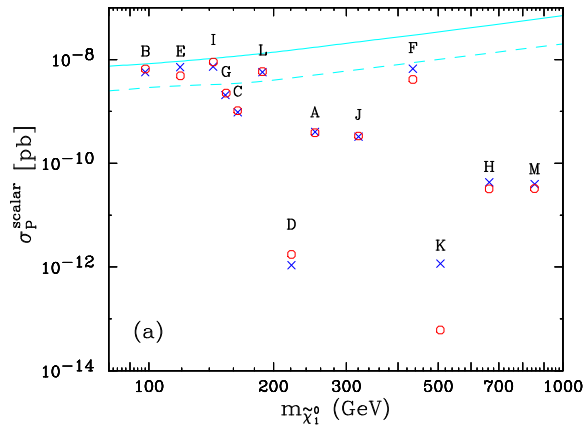}}
\caption{
The IceCube-friendly scenarios in Fig.~\ref{fig:Sun} tend also to
predict rates for the scattering of dark-matter particles that may
be accessible to direct detection.
\label{fig:direct}}
\end{figure}

As seen in Fig.~\ref{fig:direct}, these IceCube-friendly scenarios predict
that a signal should be seen in the GENIUS detector now under
construction~\cite{GENIUS}. We are all on tenterhooks to see whether some
supersymmetric
interpretation of the IceCube dark-matter signal will be confirmed. It
will also be vital to check whether the strength of this signal is
compatible with the LHC signal for sparticle production and the universal
gravity-mediated scenario for supersymmetry breaking assumed in
Figs.~\ref{fig:Sun} and \ref{fig:direct}.

\section{Neutrino Masses and Oscillations}

Back at the first $\cos \phi$ meeting, we were glimpsing an emerging
default option for neutrino phenomenology. We were becoming
convinced~\cite{SK,SNO} that there were no light sterile neutrinos besides
the the three active species confirmed by LEP. Theorists expected their
masses to be hierarchical, but there was no experimental evidence for this
hypothesis.  The atmospheric and solar experiments had told us that the
corresponding mixing angles were near maximal, and suggested that
atmospheric $\nu_\mu \to \nu_\tau$ mainly.  We had only the Chooz and
Super-Kamiokande upper limits on $\theta_{13}$. Theorists expected the
light-neutrino masses to be mainly Majorana in nature. It is also worth
remembering that no experiment had actually confirmed a neutrino
oscillation pattern, and some die-hards still advocated decay
interpretations of the data, although theorists expected neutrino
lifetimes much longer than the age of the Universe.

The neutrino-oscillation data raised almost as many questions as they
answered. Could we really exclude light sterile neutrinos, as would
certainly be required if the LSND claim were to be confirmed by MiniBooNE?
Could we really exclude degenerate neutrino masses, or an inverse mass
hierarchy? Although the LMA solution for solar-neutrino oscillations was
strongly favoured after the first SNO neutral-current measurements, some
die-hards were still clinging to the LOW and/or SMA solutions. Would
$\tau$ production be observed, and would $\theta_{13}$ be accessible to
the first generation of long-baseline neutrino experiments, a question
vital for the detectability of CP violation in neutrino oscillations?
Fundamentally, could the Majorana nature of neutrino masses be confirmed
by neutrinoless double-$\beta$ decay experiments, and/or the distinctive
ocillation pattern be observed in long-baseline experiments?

Some of our questions were answered soon after $\cos^1 \phi$, and others
more recently. The KamLAND experiment soon provided ample confirmation of
the LMA solar solution~\cite{KamLAND}, and then MiniBooNE failed to
confirm the LSND signal~\cite{MiniBooNE}. An unplanned bonus was the
observation of supernova 2007b. Detailed measurements of its neutrinos, in
particular in the reconstructed Super-Kamiokande detector, confirmed the
favoured oscillation scenario. The MINOS experiment~\cite{MINOS} duly
observed the expected oscillation pattern in charged-current reactions,
and the Gran Sasso experiments have observed $\tau$ production in the CERN
beam~\cite{OPERA+ICARUS}. Most interesting has been the saga of
$\theta_{13}$. The combination of MINOS, ICARUS and OPERA found a hint
that $\sin^2 2 \theta_{13} \sim {\rm few} \times 10^{-2}$. This indication
was
subsequently confirmed by the rejuvenated Super-Kamiokande experiment
working in the first-generation off-axis JHF neutrino beam~\cite{SKpage}.

\begin{figure}[th]
\centerline{\epsfxsize=3in\epsfbox{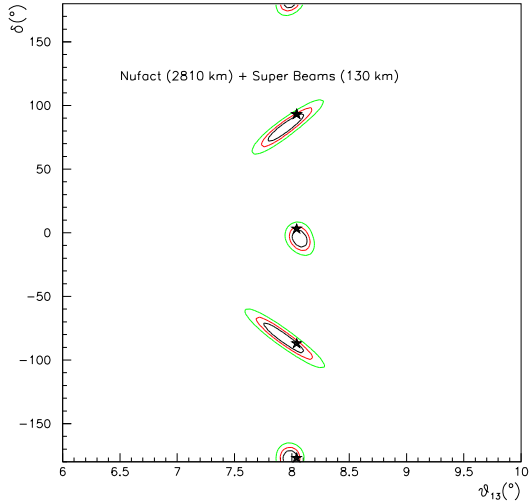}}
\caption{
A combination of the measurements of the CP-violating neutrino phase by
the superbeam and the neutrino factory will enable a precise determination
of $\delta$ to be made.
\label{fig:both}}
\end{figure}

Thus, all the elements were in place for the recent approval of a
fully-fledged neutrino factory. As you know, this should be able to answer
our remaining questions about neutrino masses and oscillations, namely the
hierarchy, degeneracy or inverse hierarchy of the mass spectrum, and the
magnitude of the CP-violating phase $\delta$. It will certainly pin down
more precisely
the magnitude of $\theta_{13}$ and measure the sign of $\Delta m_{23}^2$ 
via matter effects on the long-baseline beam. 
The best measurement of the CP-violating phase $\delta$ will be possible
via the $T$-odd 
asymmetry:
\begin{eqnarray}
P(\nu_e \to \nu_\mu) &-& P({\bar \nu_e} \to {\bar \nu_\mu}) =
16 s_{12} c_{12} s_{13} c_{13}^2 s_{23} c_{23} sin \delta \nonumber
\\
&\times&
\sin \left( {\delta m_{12}^2 \over 4 E} L \right) 
\sin \left( {\delta m_{13}^2 \over 4 E} L \right) 
\sin \left( {\delta m_{23}^2 \over 4 E} L \right).
\end{eqnarray}
at the neutrino factory. The upgraded
JHF beam and the Hyper-Kamiokande detector now under construction also have a 
good chance of making a preliminary measurement, and the combination of
the two will be very useful for resolving ambiguities, as seen in
Fig.~\ref{fig:both}~\cite{Gavela}.

In addition to its `core business' of long-baseline neutrino oscillation 
physics, the intense proton source of the neutrino factory will provide 
unique opportunities for other physics, including that using stopped or 
slow muons, which has been given considerable impetus by the recent 
results from PSI and BNL.

\section{Discovery of Lepton Flavour Violation}

Over the last decades of the 20th century, searches for the violation of
charged-lepton flavours made steady but undramatic progress in pushing
down the upper limits on observables such as $\mu \to e \gamma$ and $\mu
\to e$ conversion on nuclei. This unspectacular hard work paid off in a
big way with the recent discoveries of these two reactions at
PSI and BNL~\cite{PSI+MECO}, respectively. The two measurements are
very consistent, with the rate for $\mu \to e \gamma$ measured by MECO at
BNL being a few per-mille of the branching ratio for $\mu \to e \gamma$,
as expected in many models.

The violation of charged-lepton numbers was only to be expected at some 
level, once neutrino oscillations had been observed. However, the rates 
for such processes would have been negligible if there were no other 
low-energy particles beyond those in the Standard Model. The apparent 
observation of low-energy supersymmetry provides a good candidate for the 
new low-energy physics that enhances charged-lepton flavour violation. 
The minimal seesaw model for neutrino masses contains the following 
terms:
\begin{equation}
{L}_\nu \; = \; \left( Y_\nu \right)_{ij} (\nu, L)_i N_j H \; + {1 
\over 2} N_i {M}_{ij} N_j
\end{equation}
where the $N_i$ are three heavy singlet neutrino fields. The total number 
of physical parameters in this model is 18~\cite{EHLR}. Nine of these are
measurable 
in low-energy neutrino physics, namely 3 neutrino eigenmasses $m_\nu$, 3 
real mixing angles $\theta_{12, 23, 13}$, and 3 CP-violating phases - the 
oscillation phase $\delta$ and two Majorana phases $\phi_{1,2}$.  The 
remaining 9 parameters are needed to describe the heavy-neutrino sector, 
and again include 3 heavy eigenmasses $M_\nu$, 3 real mixing angles 
$\alpha$ and 3 CP-violating phases $\beta$. This minimal seesaw model 
therefore contains a total of 6 CP-violating phases, and the extra phases 
$\alpha$ control the rate for leptogenesis~\cite{FY}, the favoured
explanation for the origin if the baryon number of the Universe.
 
\begin{figure}[th]
\centerline{\epsfxsize=3in\epsfbox{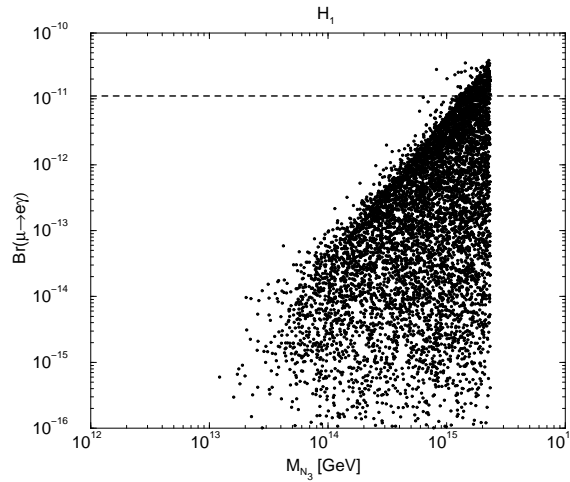}}
\caption{
The recent discovery of $\mu \to e \gamma$ decay by MEG at PSI and the
related
discovery of $\mu \to e$ conversion on a heavy nucleus by the MECO
Collaboration can easily be accomodated in the minimal
supersymmetric seesaw model.
\label{fig:muegamma}}
\end{figure}

In the minimal supersymmetric extension of this minimal seesaw model, the 
soft supersymmetry-breaking parameters are renormalized by the Dirac 
coupling matrix $Y_\nu$:
\begin{eqnarray}
\delta m^2_{\tilde \ell} & = & - {1 \over 8 \pi^2} ( 3 m_0^2 + A_0^2) 
(Y^\dagger_\nu Y_\nu)_{ij} {\rm ln} \left( {m_{GUT} \over m_{N_i}}
\right), \\
\delta A_{\ell} & = & - {3 \over 8 \pi^2} Y_{\ell_i} 
(Y^\dagger_\nu Y_\nu)_{ij} {\rm ln} \left( {m_{GUT} \over m_{N_i}}
\right).
\end{eqnarray}
Non-diagonality in the neutrino Dirac couplings $Y_\nu$ in the mass
eigenstate basis for the charged leptons induces lepton flavour violation 
in the slepton and sneutrino mass matrices, which depends on the mixing 
angles $\alpha$ and the CP-violating phases $\beta$ that are not 
observable in low-energy interactions. This mechanism is certainly able
to accommodate the MEG and MECO data, as seen in
Fig.~\ref{fig:muegamma}~\cite{EHRS}. It is certainly possible, even 
likely, that there are other sources of lepton flavour violation at the 
string and/or GUT scales, but this supersymmetric seesaw mechanism offers 
a `Standard Model' to be confronted with the MEG, MECO and other data on 
the violation of charged-lepton flavours and CP.

\begin{figure}[th]
\centerline{\epsfxsize=3in\epsfbox{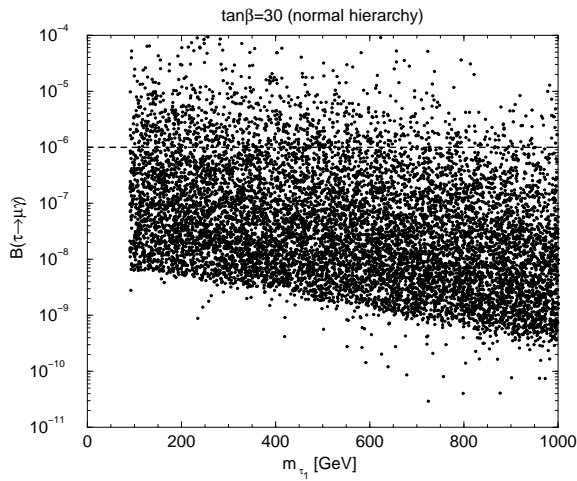}}
\caption{
The discoveries of $\mu \to e$ processes by the MEG and MECO experiments
increase the motivation that $\tau \to \mu \gamma$ may occur at a rate
close the present experimental upper limit, as rumoured from
the LHC and the $B$ factories.
\label{fig:taumugamma}}
\end{figure}

Examples of such processes include $\tau \to \mu \gamma, \tau \to e
\gamma, \mu \to 3 e, \tau \to 3 \ell$ and the electric dipole moments of
the electron and muon, which could have interesting rates in the minimal
supersymmetric seesaw model, as seen in
Fig.~\ref{fig:taumugamma}~\cite{EHRS}. Rumours have begun to circulate of
interesting rare $\tau$ decays at the LHC, and also at the $B$ factories
following their recently completed luminosity upgrades. Another
interesting place to look for lepton flavour violation is in sparticle
decays at the LHC or the $e^+ e^-$ linear collider. The suppression of
rare $\mu (\tau)$ decays is in large part due to loop factors and the
relatively large ratio of slepton over lepton masses, and relatively large
branching ratios for $\chi_2 \to \chi \ell^+ \ell^{- \prime}$ are quite
possible~\cite{CEGLR}. Here again, some interesting hints have been
emerging from the LHC, but we may have to wait for $\cos^{11} \phi$ and/or the LHC
luminosity upgrade before these are confirmed.

\section{Space-Time Foam}

We have known for over a decade now, thanks to the early experiments on
the cosmic microwave background radiation, that the geometry of the
Universe is flat on large distance scales. However, we have long expected
that it should exhibit large quantum fluctuations on small distance and
time scales:

\begin{equation}
\Delta E, \Delta \chi = {O}(1) \; \; {\rm in} \; \;
\Delta x, \Delta t = {O}(1),
\label{stfoam}
\end{equation}
where the energy $E$, distance $L$ and time $t$ are measured in Planck 
units $\sim 10^{19}$~GeV, $10^{-33}$~cm and $10^{-43}$~s, respectively, 
and $\chi$ is a generic measure of space-time topology. The big question 
has been whether there could be any observable consequences of this 
`space-time foam' (\ref{stfoam})? Most theorists have thought this must
surely be 
impossible, but some disreputable characters have suggested that there
might be observable loss 
of information across microscopic event horizons~\cite{H,EHNS}, modifying
the conventional superposition rules of quantum mechanics, and also 
that the `recoil' of the vacuum during the passage of an 
energetic particle might modify its apparent velocity~\cite{AEMN}:
\begin{equation}
c(E) \; = \; c_0 ( 1 - {E \over M} + \cdots),
\label{Lorentz}
\end{equation}
where $c_0$ is the low-energy velocity of light and and one might expect 
that the effective quantum-gravity scale $M \sim M_P$. This 
possibility first arose in the context of a string-inspired model of 
space-time foam~\cite{EMN}, but subsequently found possible in the more
traditional loop approach to quantum gravity.
Microscopic tests of quantum mechanics have not advanced much since the
old days of CPLEAR~\cite{CPLEAR}, over a decade ago, but there have been
some
interesting recent developments related to the suggestion (\ref{Lorentz}).

\begin{figure}[th]
\centerline{\rotatebox{-90}{\epsfxsize=1.5in\epsfbox{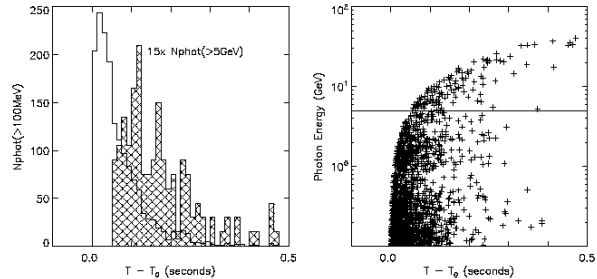}}}
\caption{
The dispersions in arrival times of energetic photons from gamma-ray
bursters observed by the GLAST detector may be similar to sumulations
incorporating an energy-dependent velocity of light.
\label{fig:GLAST}}
\end{figure}

Some of the best opportunities for probing this possibility are provided 
by astrophysical sources~\cite{AEMNS}: these offer long light propagation
times $t = 
D/c$ and hence relatively large time delays $\delta t \sim (D / c^2) (E / 
M)$. Astrophysical sources with short intrinsic fluctuation time scales 
$\Delta t$ yield the best figures of merit for probing the quantum-gravity 
scale $M$:
\begin{equation}
M \sim {E \cdot D \over \Delta t}.
\end{equation}
Examples of interesting astrophysical sources include gamma-ray bursters 
(GRBs), pulsars and active galactic nuclei (AGNs). Back at the beginning 
of the past decade, the latter already yielded limits~\cite{GRBs}
\begin{equation}
M > 2 \times 10^{16}~{\rm GeV},
\end{equation}
and the GLAST collaboration had made preliminary estimates of their 
sensitivity to this possible quantum-gravity effect, as seen in 
Fig.~\ref{fig:GLAST}~\cite{Norris}. Recently, we have started hearing
rumours that GLAST
observations of some GRBs exhibit unexplained time-lags~\cite{GLAST}. It
is in 
principle possible to distinguish propagation effects from time-lags at 
the sources, by looking for a correlation with distance $D$ for GRBs with 
known redshifts $z$~\cite{GRBs}, and we look forward to hearing the
results of such an 
analysis.

\section{Probing the String-Particle Phase Transition}

With the release of the maps of the cosmic microwave background (CMB) sky
by the Planck satellite~\cite{Planck}, experimental tests of the
string-particle phase transition have begun in earnest. Over a decade
ago, it
started dawning on cosmologists that the CMB could be sensitive to
trans-Planckian physics~\cite{transPlanck}, then string theorists got in
on the action, and the rest, as they say, is history. Non-perturbative
string phenomenology is now in full swing, with attempts to match the
Planck polarization data, the apparent deviations from scale invariance
and the indications of non-Gaussian behaviour of the CMB fluctuations.

\begin{figure}[th]
\centerline{\epsfxsize=3.5in\epsfbox{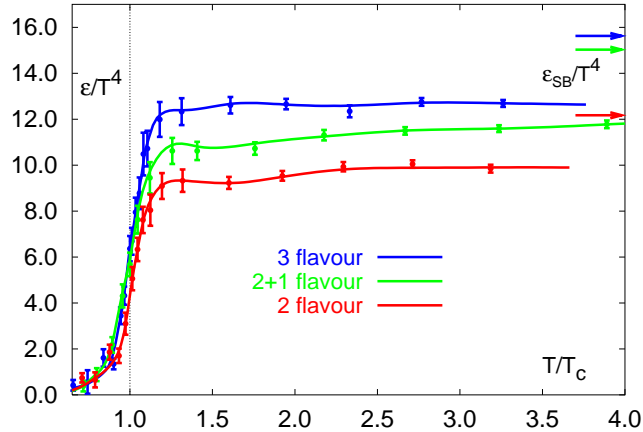}}
\caption{
Recent simulations of the string-particle transition using Petaflop
computer Grids indicate a behaviour analogous to that found for the
quark-hadron phase transition, and have observable signatures
in the cosmic microwave background radiation.
\label{fig:QH}}
\end{figure}

One the most interesting recent calculational developments has been the
debut of lattice simulations of the string-particle transition and the
emergence of space-time using the latest Petaflop computer Grids. We now
know that there are many similarities between the behaviours of bulk
quantities across this transition, such as the metric, and those in the
corresponding quark-hadron transition, depicted in
Fig.~\ref{fig:QH}~\cite{Satz}, which depend on the details of the string
compactification.

\section{Theoretical Advances}

Who would have imagined back at the first $\cos \phi$ conference that we
would be measuring directly the emergence of space-time from the string
soup, and that we would be well on the way to calculating it? And who
would have believed that we would now know how to calculate the
present-day vacuum energy in terms of the soft supersymmetry-breaking
parameters and the string vacuum moduli, as now seems `trivial'?

These and other remarkable developments have been made possible by the 
impressive theoretical advances since the first $\cos \phi$ conference. 
Thinking back to that meeting, the list of problems still unsolved then
seems almost comical:

\begin{itemize}

\item{The mechanism of supersymmetry breaking,}

\item{Vacuum energy,}

\item{How to fix the vacuum moduli,}

\item{The cosmological gravitino problem,}

\item{The supersymmetric CP-violating phases,}

\item{The sizes of the extra dimensions,}

\end{itemize}
\noindent
and many more. This audience would be bored if I described how all these 
problems have been solved, and the margin of this proceedings contribution
is any case too small. Here it suffices to say that 
$ABCDEFGHIJKLMNOPQRSTUVWXYZ$ 
theory does indeed solve them, impossible as this might have seemed back 
before the $N > 2$ string revolutions, when our only 
theoretical tools were $M$, $F$ and $K$ theory.

\section*{Acknowledgements}

I thank Alon Faraggi and Steve Abel for their foresight back in 2002,
asking me to undertake this exercise in hindsight. No doubt my memory has
become hazy on some points, and I apologize in advance to any 2012 readers
for its comical lapses. They will probably consider many other discoveries
to have been more exciting. Finally, I think the Yukawa Institute in Tokyo
for its hospitality while writing up this load of crystal ball.

\end{document}